\documentclass{aa}
\usepackage{psfig}
\usepackage{epsfig}
\def \src {X\thinspace 1624-490}
\def \th {\thinspace}
\def \sax {{\it BeppoSAX}}

\def \nh {N${\rm _H}$}

\def \hcm {\hbox {\ifmmode $ atoms cm$^{-2}\else atoms cm$^{-2}$\fi}}
\def \arcmin {\hbox{$^\prime$}}
\def \arcsec {\hbox{$^{\prime\prime}$}}

\def\approxgt{\mathrel{\hbox{\rlap{\lower.55ex \hbox {$\sim$}}
        \kern-.3em \raise.4ex \hbox{$>$}}}}

\def\approxlt{\mathrel{\hbox{\rlap{\lower.55ex \hbox {$\sim$}}
        \kern-.3em \raise.4ex \hbox{$<$}}}}
\begin{document}

   \thesaurus{6(13.25.5; 
               08.09.2: X\th 1624$-$490;  
               08.14.1;  
               08.02.1;  
               02.01.2;  
               09.04.1)} 
   \title{A study of the dipping low mass X-ray binary X\th 1624-490 from the
	  broadband BeppoSAX observation}

   \author{M. Ba\l ucinska-Church\inst{1} 
	\and P.J. Humphrey\inst{1}
	\and M.J. Church\inst{1}
        \and A.N. Parmar\inst{2}}

   \offprints{M. Ba\l uci\'nska-Church (mbc@star.sr.bham.ac.uk)}
   \institute{School of Physics and Astronomy, University of Birmingham,
              Birmingham B15 2TT\\
              email: mbc@star.sr.bham.ac.uk
         \and
           Astrophysics Division, Space Science Department of ESA, ESTEC
		Postbus 299, NL-2200 AG Noordwijk, The Netherlands\\
             email: aparmar@astro.estec.esa.nl}

   \date{Received 28 February 2000; Accepted 24 May 2000 }

\maketitle

\markboth{M. Ba\l uci\'nska-Church et al.}{A study of X\th 1624-490}

   \maketitle

\begin{abstract}
We present results of a study of the luminous dipping low mass X-ray
binary \src\ made using {\it BeppoSAX}. An interval of deep and rapidly varying dipping was included 
in the observation. The radial intensity 
profile of the source obtained using the MECS instruments revealed excesses in intensity 
above the instrument point spread function below $\sim $5 keV demonstrating 
the presence of a dust-scattering halo. From modelling of the radial profile in several 
energy bands, halo intensity fractions rising to 30\% in the lowest band 2.5--3.5 keV
were obtained. From these data, the optical 
depth to dust scattering at 1 keV was found to be $\rm {2.4\pm 0.4}$.
The non-dip spectrum of \src\ in the energy band 
1--100 keV is shown to be well-described by the emission model 
consisting of point-like blackbody radiation assumed to be from the neutron star 
plus extended Comptonized emission from an ADC. The blackbody 
temperature was $\rm {1.31\pm 0.07}$ keV and the Comptonized emission had photon 
power law index $\rm {2.0^{+0.5}_{-0.7}}$ and cut-off energy $\sim $12 keV. The spectra of 
several dip levels were shown to contain an unabsorbed component below
5 keV. Good fits to the dip spectra were obtained by allowing the Comptonized 
emission to be progressively covered by an extended absorber while the blackbody was rapidly 
absorbed and a constant halo component accounted for dust scattering
into the line-of-sight. It is shown that the unabsorbed component consists 
of the uncovered part of the Comptonized emission plus a halo
contribution which in deepest dipping dominates the spectrum below 4.5 keV. From the dip ingress 
time, we have derived a diameter of the extended Comptonized emission region of 
$\rm {5.3\pm 0.8\times 10^{10}}$ cm, consistent with a hot, X-ray emitting corona 
extending to $\sim $50\% of the accretion disk radius.
The source luminosity for a distance of 15 kpc is $\rm
{7.3\times 10^{37}}$ erg
s$^{-1}$, an appreciable fraction of the Eddington limit making \src\ the most luminous dipping 
LMXB. The half-height of the blackbody emitting region on the neutron star of 
$\rm {6.8\pm 1.8}$ km agrees with the half-height of the radiatively supported inner accretion disk
of $\rm {6.3\pm 2.9}$ km, which together with  similar agreement recently obtained for 13 
other LMXB strongly supports the identification of the neutron star as the origin of the blackbody
emission in LMXB. Finally, from {\it RXTE} ASM data, we derive an
improved orbital period of 20.87$\pm $0.01 hr.
\end{abstract}


\section{Introduction}
\src\ is one of the most unusual members of the class of dipping Low Mass X-ray Binary
(LMXB) sources exhibiting periodic dips in X-ray intensity. It is generally accepted
that dipping is due to absorption in the bulge in the outer accretion disk where the
accretion flow from the companion inpacts (White \& Swank 1982). \src\ has the longest
orbital period of the dipping sources at 21$\pm $2 hr (Watson et al.
1985), but the period has not been refined since this determination
from {\it Exosat}. Dipping is deep, $\sim$ 75\% in the band 1--10 keV,
and the source also exhibits strong flaring in which the X-ray flux can increase by 
30\% over timescales of a few thousand seconds (Church \& Ba\l uci\'nska-Church 1995). 

The depth, duration and spectral evolution in dipping vary considerably from source
to source. However, spectral analysis of dipping sources provides information 
not available in non-dip sources. For example, spectral models are more
strongly constrained by having to fit non-dip and dip data, thus showing clearly
the nature of the emission. In addition, the dip ingress and egress times
can be used to obtain the sizes of the extended emission regions when the
absorber has larger angular extent than the source regions, thus providing 
information on the geometry of the accretion disk corona (ADC) (e.g. Church et al. 1997). Although
the fundamental question of the nature of the emission regions in LMXB is
controversial, a unifying model has been proposed for the dipping class (Church
\& Ba\l uci\'nska-Church 1995) which is able to explain the widely disparate 
behaviour of individual sources. The emission regions consist of the point source 
neutron star producing blackbody emission, and an extended Comptonizing region, 
probably the ADC. It has now been shown that this model not only can
explain well the spectra of {\it all} dipping LMXB (Church et al. in preparation),
but can also explain the spectra of all the Z-track and Atoll sources investigated in a
recent {\it ASCA} survey (Church \& Ba\l uci\'nska-Church 2000).
In a number of dipping sources, the nature of dipping
was not understood because of the clear presence of an unabsorbed part of the non-dip
spectrum at all levels of dipping. In these sources, dip spectra could be modelled
by dividing the non-dip spectral form into two parts: one absorbed and the other
unabsorbed but with strongly decreasing normalization (Parmar et al. 1986; Courvoisier 
et al. 1986; Smale et al. 1992), `absorbed plus unabsorbed' modelling. It has been
difficult however, to explain these normalization changes convincingly. More recently, 
an alternative explanation has been suggested in which an extended absorber moves 
across the extended and point-like emission regions, providing a natural explanation 
for the unabsorbed component as the uncovered emission which becomes smaller as dipping 
proceeds. This Progressive Covering model has explained spectral evolution during
dipping in XB\th 1916-053, XB\th 0748-676 and XB\th 1323-619 (Church
et al.  1997, 1998a,b; Ba\l uci\'nska-Church et al. 1999).

\src\ was previously observed in a long {\it Exosat} observation of 220 ks and with {\it Ginga}
(Jones \& Watson 1989). The {\it Exosat} ME data revealed an apparently 
stable lower level of dipping supporting the presence of two emission components, one of 
which was totally absorbed in deep dipping. A blackbody
plus bremsstrahlung model was used to parameterize the spectra (Jones \& Watson 1989).
Church \& Ba\l uci\'nska-Church (1995; hereafter CBC95)
showed that the light curve at higher energies ($>$ 5 keV)
was dominated by flaring which can strongly modify the spectrum and make the spectral
investigation of dipping difficult. By selecting sections of non-dip and dip data
without apparent flaring, the above unifying blackbody plus
Comptonization model was found to fit the data well
showing that in deep dipping the blackbody component was totally absorbed, and that
the Comptonized component was relatively little absorbed. However, with only one
non-dip and one deep dip spectrum it was not possible to determine 
the extent of absorption of the Comptonized component. The Galactic
column density of \src\ is very high $\sim \rm {8\times 10^{22}}$ atom
cm$^{-2}$ so that a dust scattered halo of the source is expected, and
Angelini et al. (1997) demonstrated an excess of the surface brightness 
above the point spread function in {\it ASCA} GIS. 

In the present paper, we present a detailed study of dipping in X\th 1624-490 and of the 
effects of dust scattering in this source made using {\it BeppoSAX}. With the very broad band, 
we have been able for the first time to obtain reliably parameters of the Comptonized 
emission. We also present a determination of the orbital period of \src\ from 
the {\it RXTE} ASM.

\section{Observations}
\label{sec:observations}

Data from the Low-Energy Concentrator Spectrometer (LECS;
0.1--10~keV; Parmar et al. 1997), Medium-Energy Concentrator
Spectrometer (MECS; 1.3--10~keV; Boella et al. 1997),
High Pressure Gas Scintillation Proportional Counter
(HPGSPC; 5--120~keV; Manzo et al. 1997) and the Phoswich
Detection System (PDS; 15--300~keV; Frontera et al. 1997) on-board \sax\
are presented. All these instruments are coaligned and collectively referred
to as the Narrow Field Instruments, or NFI.
The MECS consists of three identical grazing incidence
telescopes with imaging gas scintillation proportional counters in
their focal planes, however prior to the observation of \src\ one of
the detectors had failed. The LECS uses an identical concentrator system as
the MECS, but utilizes an ultra-thin entrance window and
a driftless configuration to extend the low-energy response to
0.1~keV. The non-imaging HPGSPC consists of a single unit with a collimator
that is alternatively rocked on- and off-source to monitor the
background spectrum. The non-imaging
PDS consists of four independent units arranged in pairs each having a
separate collimator. Each collimator can be alternatively
rocked on- and off-source to monitor the background.

The region of sky containing \src\ was observed by \sax\
between 1999 August 11 00:47 and August 11 18:22~UTC.
Good data were selected from intervals when the elevation angle
above the Earth's limb was $>$$4^{\circ}$ 
for LECS and MECS, $>$$5^{\circ}$ for the HPGSPC and $>$$10^{\circ}$
for the PDS and when the instrument
configurations were nominal, using the SAXDAS 1.3.0 data analysis package.
The standard collimator dwell time of 96~s for each on- and
off-source position was used, together with rocking angles of 180\arcmin\
and 210\arcmin\ for the HPGSPC and PDS, respectively. The exposures
in the LECS, MECS, HPGSPC, and PDS instruments are 16~ks, 34~ks,
17~ks, and 16~ks, respectively. LECS and MECS data were
extracted centered on the position of \src\ using radii of 8\arcmin\ and
4\arcmin. Background subtraction in the imaging instruments
was performed using standard files scaled appropriately
to match the background level in the image, this being required because of the location 
of the source in the Galactic plane, but is not critical for such a
bright source. Background subtraction in the non-imaging instruments was
carried out using data from the offset intervals.

\begin{figure*}[!ht]
\epsfxsize=110 mm
\begin{center}
\leavevmode
\epsffile{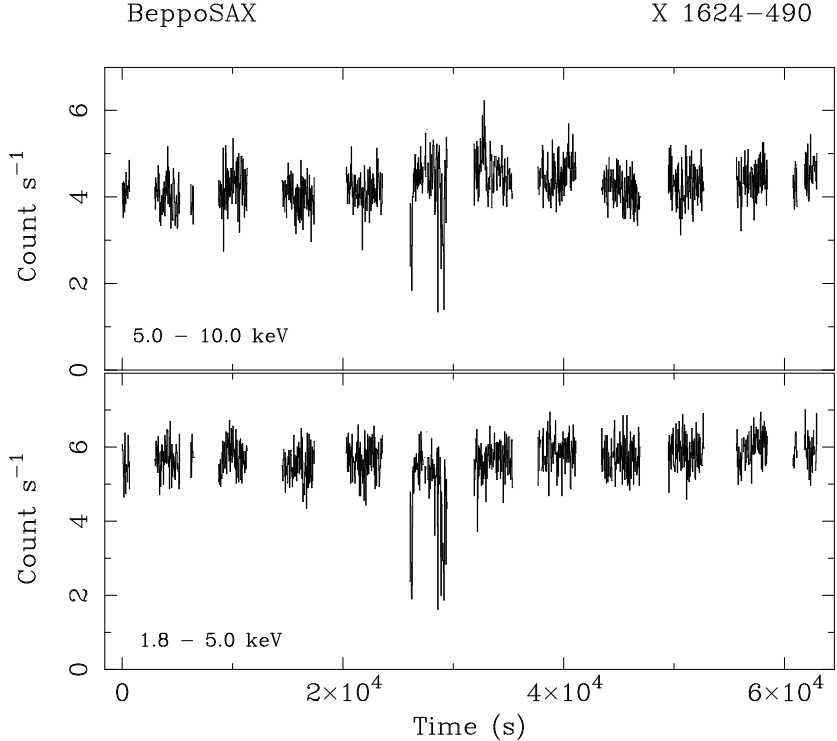}
\end{center}
\caption{MECS light curves of X\th 1624-490 in two energy bands:
1.8--5.0 and 5.0--10.0 keV with 32 s binning}
\label{}
\end{figure*}
\section{Results}

\subsection {The X-ray lightcurve}

Figure 1 shows the background-subtracted MECS lightcurves of \src\ with 
a binning of 32~s in two energy bands: 1.8--5.0 keV and 5.0--10.0 keV.
Observation of dipping is difficult in this source with a long
orbital period if the total observation time is to be kept within limits. In this case
the observation was carried out on the basis of a dipping ephemeris we
obtained from the {\it RXTE} ASM, aiming for the dipping to occur at the centre of
the observation. 
Dipping was detected as seen in Fig. 1, lasting between 3.6 and 8.4 ks with parts being missed
in data gaps. The total count rate was $\sim $11 count s$^{-1}$.
Data in the high energy band reveal little flaring which was clearly
seen in the {\it Exosat} ME as increases in count rate above 5 keV
lasting several thousand seconds (CBC95), except for an increase in 
count rate of 20\% at $\sim $32 ks lasting 1.4 ks. 
We detected little effect in the spectral fitting that 
might be due to flaring. 
An expanded view of the dipping in the band 1.8--10 keV is shown in
Fig.~2 revealing strong variability on timescales of $\sim $32 s proving 
``blobbiness'' in the absorber. Dips are not strongly evident in the HPGSPC and 
PDS lightcurves due to the comparatively reduced signal-to-noise ratio in these instruments 
and the energy dependence of the dipping.

It can be seen from Fig.~1 having 32 s binning that dipping is not 100\% deep 
in either energy band
unlike, for example, XB\th 1916-053 in which dipping often reaches 100\% depth 
at all energies below 10 keV (Church et al. 1997). However, because of
the fast dip ingress and egress times, 32 s binning is not short
enough to reveal the true depth. Tests showed that 8 s binning was
adequately short, in which case the depth of deepest dipping was $\rm {83\pm 5}$\%
and $\rm {78\pm 5}$\% 
\begin{figure}[!ht]
\epsfxsize=80 mm
\begin{center}
\leavevmode
\epsffile{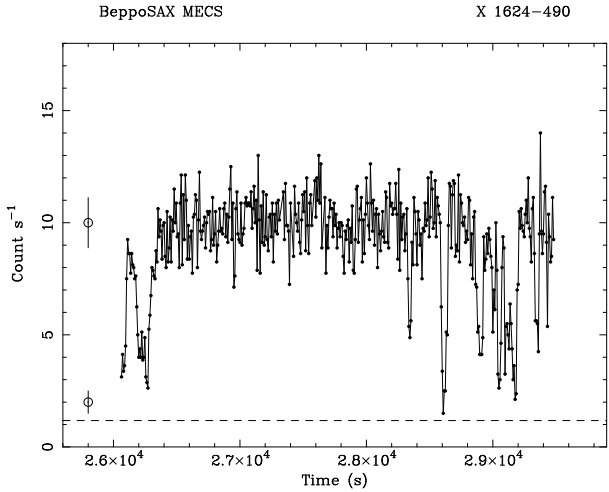}
\end{center}
\caption{Expanded view of the light curve in the energy range
1.8--10 keV with 8s binning. For clarity, Poisson errors are not shown
on every point but are indicated separately for two intensity levels.
The dotted line shows the expected contribution in this band of the
dust-scattered halo from our radial modelling (Sect. 3.3)}
\label{}
\end{figure}
in the bands 1.8--5.0 keV and 5.0--10.0 keV
respectively. In Fig.~3 we show the results of determining the depth
of dipping in the energy bands 2.5--3.5 keV, 3.5--4.5 keV, 4.5--5.5
keV, 5.5--6.5 keV and 6.5--7.5 keV from MECS data with 8 s binning.
Results are shown for data extracted in the standard radius of
4\arcmin\, and also for extraction radii of 2\arcmin\ and 15$^\prime$.
For 15\arcmin\ extraction, it was necessary to subtract background
to avoid errors in the depth.
It was not possible to obtain results outside the overall band of
2.5--7.5 keV because of poor statistics.
Although there is no dramatic change in depth with energy, 
\begin{figure}[!h]
\epsfxsize=80 mm
\begin{center}
\leavevmode
\epsffile{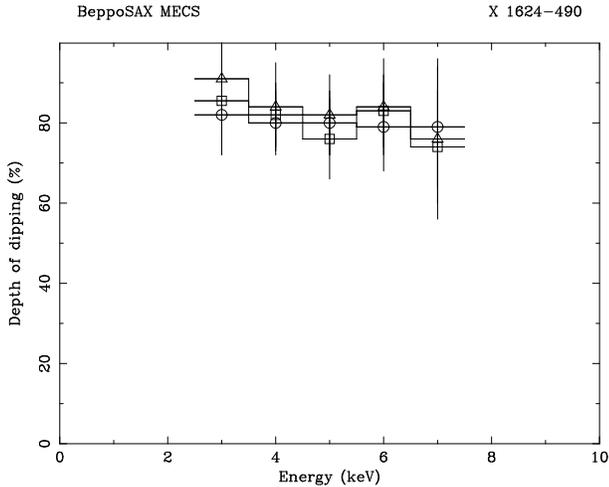}
\end{center}
\caption{Depth of dipping in X\th 1624-490 in several energy bands in
the energy range
2--10 kev from MECS data for data extraction from 15\arcmin\ radius
(squares), 4\arcmin\ radius (circles) and 2\arcmin\ radius (triangles)}
\label{}
\end{figure}
there is evidence for the depth increasing at low energies, particularly for
2\arcmin\ extraction with 90\% deep dipping at 2.5--3.5 keV. 
There is also evidence that dipping was deeper with 2\arcmin\
extraction, but no evidence for differences between 4\arcmin\ and
15\arcmin\ extraction.

Thus dipping is $\sim $ 80\% deep over much of the 1--10 keV band, but
is not actually proven to be 100\% at any energy. There can be two reasons for 
this. Because the Galactic column density in \src\ appears to be high,
dust scattering is expected to be important and below 5 keV this can result 
in a non-zero intensity level in deepest dipping. Above $\sim $5 keV, dust scattering will not
be important, and non-zero dip intensity can be due to
incomplete absorption of one emission component.
The increase in depth of dipping in 2\arcmin\
extraction is consistent with a decreased amount of dust-scattered halo.
The column density of X\th 1624-490 from spectral fitting is high and so we 
have investigated possible effects of dust scattering. This is discussed 
in detail in Sect. 3.3.

\begin{figure}[!t]
\epsfxsize=80 mm
\begin{center}
\leavevmode
\epsffile{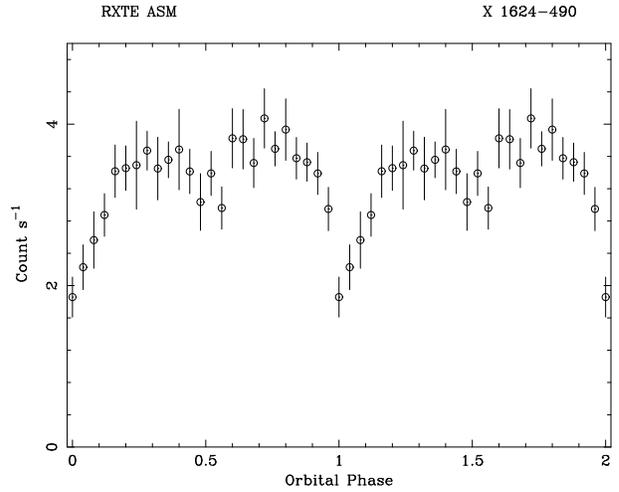}
\end{center}
\caption{Folded light curve of {\it RXTE} ASM data produced from 32 days
of data folded on a period of 20.87 hr}
\label{}
\end{figure}

\subsection{The orbital period}

We cannot carry out folding analysis of the present \sax\
X-ray light curve to determine 
the orbital period; however, we have examined the {\it RXTE} ASM data on this source 
to determine the period more accurately which previously was only poorly known
at 21$\pm $2 hr (Watson et al. 1985).
Data were used between 1996, Feb. 20 and 2000, Jan. 18 and a period search 
carried out using standard (XRONOS) power spectrum software 
spanning the period range 2.5 -- $\rm {4\times 10^4}$ hr. 
Peaks were seen with high 
significance corresponding to periods of 20.87$\pm \rm {0.01}$ hr and 
10.438$\pm \rm {0.001}$ hr, the significance of each of these peaks
corresponding to a probability $>>$ 99.8\% of being real. 
32 days of data folded on the period of 20.87 hr are shown 
in Fig. 4, in which dipping is clearly seen to repeat at 20.87 hr with
interdipping also seen between the main dips. This is the first
detection of interdipping in this source. 
In the four years of ASM data analysed, the depth
of dipping investigated in 32 day sections of data varies, and in particular, 
the depth of interdipping varies from being as deep as the main dipping
to being undetectable. There is some evidence that this pattern repeats
on a timescale of $\sim $70 days, but this cannot yet be identified 
definitely as a new periodicity in the system. The 10.438 hr periodicity
cannot be the orbital period for several reasons: firstly this
periodicity is at times not present in the ASM data when the 20.87 hr
periodicity is present. Additionally, Fig. 4 shows dipping at 10 hr
from the main dips is weak, entirely consistent with interdipping. 
If the orbital period were 10.438 hr, then a strong peak at 21 hr would not be
seen in the power spectrum. Finally, the {\it Exosat} ME lightcurve
(Jones \& Watson 1989; Church \& Ba\l uci\'nska-Church 1995) allows
the possibility of a 10-hr orbital period to be almost certainly
rejected.

\subsection {The dust scattering halo}

Although the column density of X\th 1624-490 is high ($\rm {\sim 9\times 10^{22}}$ 
atom cm$^{-2}$ in the present work), implying that dust scattering
effects will be important, radio measurements have indicated a lower 
value of $\rm {\sim 2\times 10^{22}}$ atom cm$^{-2}$ 
(Dickey \& Lockman 1990; Stark et al. 1992) (although these
measurements may underestimate $\rm {N_H}$ when this is high), so that 
there is a possibility that the high value of 
$\rm {N_H}$ is due to absorption intrinsic to the source. To investigate
dust scattering, we have carried out modelling of the radial intensity distribution.
This was done using MECS data since spectral evolution during dipping
was investigated primarily with MECS. The modelling was based on the
technique used by Predehl \& Schmitt (1995; hereafter PS95) in their investigation of 25 Galactic 
sources (not including X\th 1624-490) subject to high Galactic column densities 
using the {\it Rosat} PSPC. In the case of \src\, we find excesses in the radial
profile above the point spread function (PSF) of the
instrument for radii greater than $\sim $100$^{\prime\prime}$
indicating a substantial halo contribution. Software has been 
developed to allow fitting the radial distribution including the source
contribution convolved with the PSF, a dust-scattered halo calculated
on the basis of Rayleigh-Gans scattering theory with its associated radial 
distribution function (Predehl \& Klose 1996; PS95), plus a background contribution. 
As the source lies in the Galactic Plane, the standard background files
are not applicable, and we adopted the same approach as PS95 and allowed
the background to be a free parameter. A more complete treatment
would also convolve the halo with the energy-dependent PSF.
However, we investigated the effect of this and found
that the value of the optical depth to dust scattering would only be reduced by
up to 20\%. The point spread function of Boella et al. (1997) was utilised and 
XIMAGE used to provide a radial distribution function of the data.
From the best fit of the model to the data, the halo intensity fraction {\it f$_h$} was derived,
where {\it f$_h$} is defined as the fraction of the observed intensity due to the halo, 
and may be written in terms of the observed source intensity $\rm {I_x}$ and the observed halo 
intensity $\rm {I_h}$ as \[\rm {f_h = I_{h}/(I_x\,+\,I_{h}).}\]
These procedures were tested by applying them to the bright Z-track source 
GX\th 17+2 which has been observed both with {\it Rosat} and {\it BeppoSAX}. Firstly, we
modelled the PSPC data on GX\th 17+2 and derived results similar to those of PS95. Next, we
modelled the \sax\ MECS data on this source. In the case of GX\th 17+2 only, PS95 obtained
halo fractions in 7 energy bands within the overall PSPC band, which has allowed us to compare
MECS results with their results at an energy where the MECS and the
PSPC bands overlap. In
the band 1.7--2.1 keV a good fit was obtained with a $\chi ^2$/dof of 78/74, and a halo
intensity fraction at 1.9 keV of $\rm {27\pm 2}$\%, compared
with the value of 26\% which 
\begin{figure}[!ht]
\epsfxsize=80 mm
\begin{center}
\leavevmode
\epsffile{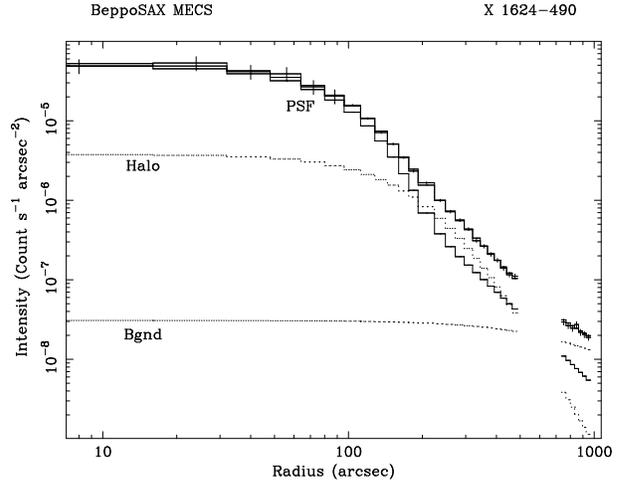}
\end{center}
\caption{Radial fitting to the MECS radial intensity distribution in the
energy band 2.5--3.5 keV. The data points, the best-fit total model, the
X-ray source component (convolved with the PSF), the halo component
and the background contribution are shown separately}
\label{}
\end{figure}
we obtain from Fig. 8 of PS95 at this energy. This good
agreement gives us confidence that the modelling was correct, and that reliable
values of {\it f$_h$} can be obtained using {\it BeppoSAX}. Next, modelling was carried out 
for X\th 1624-490.
Initially the MECS data were binned to a minimum of 20 counts per radial bin, then further 
grouping was applied with radial bins grouped together in pairs 
to reduce fluctuations. Systematic errors of 10\% were added
between 10--100\arcsec\ where the PSF is uncertain by about this amount,
and 2\% between 100-1000\arcsec\ to account for uncertainties in 
the PSF, faint, barely resolved sources in the field of view, and other effects. 
Radial bins between 500-700\arcsec\ were ignored where a detector
support structure causes an artificial reduction in the radial profile.
The fitting was done in four energy bands: 2.5--3.5, 3.5--4.5, 4.5--5.5 and 5.5--6.5 keV.
The best-fit solution is shown in Fig. 5 for the lowest band
2.5--3.5 keV in which the halo was strongest. Values of the 
halo fraction {\it f$_h$} from the best-fit models were found to be: 
$\rm {30\pm 4}$\%, $\rm {20\pm 4}$\%, $\rm {5.7\pm 2}$\% in the three lowest 
energy bands. In the band 5.5--6.5 keV, there was no evidence for any departure of
the radial profile from the PSF, and 
no improvement in the quality of fit by including a halo term in the model.
We also calculated the halo fractions of the total count from these fitting results
by integrating to a restricted outer radius of 4$^\prime$, the extraction radius 
used for MECS spectra (Sect. 3.4). The fractions became: $\rm {26\pm 4}$\%, 
$\rm {19\pm 4}$\%, $\rm {5.7\pm 2}$\% and $<$ 2\% respectively, in the four bands. 
When these fractions are compared with the depth of dipping in these
bands (Fig. 3), it can be seen that the halo cannot explain the fact
that dipping does not reach a depth of 100\%, particularly at the higher
energies. As discussed in Sect. 3.5, this must be due to incomplete absorption  
of an extended emission component. In Fig. 2 we compare the
halo contribution to the intensity (dashed line) calculated from our results for the band
1.8--10 keV with the depth of dipping.

Using this limited number of energy bands, we have compared the energy variation 
with theoretical expectations, and derived the optical depth to dust scattering 
$\tau $, evaluated at 1 keV. The reduction in source intensity is given by 
$\rm {I_x\; =\; I_0\; e^{-\tau}}$, 
from which it can be shown that the halo 
fraction and $\tau $ are related {\it via} $\rm {f_h\; =\; 1\; -\; e^{-\tau}}$, 
provided that the intensity scattered out of the line-of-sight is balanced by that
scattered into the line-of-sight as is normally assumed (e.g. Martin 1970).
Since the dust scattering cross-section is 
expected theoretically to vary as $\rm {E^{-2}}$ (Mauche \& Gorenstein
{\bf }1986), then the optical depth 
will also have this dependence, and 
\[\rm {f_h\; =\; 1\; -\; e^{ -\tau_1\; E^{-2}}}\]
where $\rm {\tau_1}$ is the optical depth at 1 keV, normally quoted.
Using this relation, we have derived a value at 1 keV of $\tau $ = 2.4$\pm \rm{0.4}$.
Within the errors on the individual values of {\it f$_h$}, the data
are consistent with an $\rm {E^{-2}}$ dependence of the cross-section.

The results show that dust scattering in the case of \src\ is large. Furthermore,
we can compare the above value of $\tau $ with values in Fig. 7 of
PS95, in which $\tau $ is plotted against the X-ray column density
for the 25 Galactic sources in this survey, and varies between zero and $\sim $1.5.
Thus in X\th 1624-490, $\tau $ is much larger.
The value of $\tau $ we obtain, and the value of
$\rm {N_H}$ from spectral fitting of $\rm {8.6\times 10^{22}}$ atom
cm$^{-2}$ may be compared with the linear $\tau $--$\rm
{N_H}$ relation obtained by PS95. This 
best relation: $\tau $ = $\rm {0.5\,N_H\,[10^{22}]-\,0.083}$ 
predicts in our case ($\tau $ = 2.4) that $\rm {N_H}$ = 
$\rm {5.0\times 10^{22}}$ atom cm$^{-2}$, compared with our fitting
result of $\rm {8.6\times 10^{22}}$ atom cm$^{-2}$ for the
non-dip spectrum. The difference between these may imply a flattening
of the relation at higher column density, or possibly intrinsic
absorption in the source.

\subsection {X-ray spectra}

It has been shown that a substantial dust scattering halo is observed
in X-rays around X\th 1624-490, and we now consider prior to presenting
spectral fitting results how spectra will be affected. In the case of
non-dip spectra, the effect is, in fact,  expected to be zero
(provided all of the halo is collected), since radiation
removed from the line-of-sight is replaced by that
scattered into the line-of-sight by scattering, both effects taking
place predominantly at lower energies, so that the net effect on the
non-dip spectrum is zero. The consequence of this is that the observed
intensities due to halo: $\rm {I_h}$ and due to the source $\rm {I_x}$
are equal to the source intensity that would have been observed
without scattering; i.e. 

\[\rm {I_0\; =\; I_x\; +\; I_h}\]
Consequently, we did not include a dust scattering term
in fitting the non-dip spectra. However, in dipping, the intensity
lost by dust scattering is proportional to the dip source intensity
while the intensity gained depends on the non-dip intensity because
of the time delays in scattered radiation compared with unscattered.
Thus, in dip data there will be an additional dust scattered
component. We have carried out spectral fitting of the MECS dip spectra
firstly without any additional component, and then secondly adding terms
to account for the effects of dust scattering both out of and into the
line-of-sight. The total halo
contains both blackbody and cut-off power law terms from the spectrum
incident to scattering.

The spectrum of non-dip emission was investigated by simultaneously
fitting data from all the \sax\ NFI. Non-dip MECS data were selected by
choosing a narrow intensity band from 9.7--10.3 count s$^{-1}$, also
removing the dip data by time filtering. 
LECS, HPGSPC and PDS data corresponding to these MECS data were selected. 
Spectral evolution in dipping was investigated using MECS data only
since the dips are not strongly seen at higher energies and there are too
few counts in the LECS to contribute meaningfully. For non-dip
data, the LECS and MECS spectra were rebinned to oversample the full width
at half maximum of the energy resolution by a factor of 3, and
additionally  LECS  data were rebinned to a minimum of 20 counts per bin
and MECS data to 40 counts per bin to allow use of the $\chi^2$ statistic.
The LECS and MECS non-dip spectra were also more heavily grouped to 
a minimum of 100 counts per bin in each case, as experience has shown that 
more stable fitting may result. However, in this case, results obtained
with the alternative groupings were in excellent agreement.
LECS data were only used between 1--4 keV and MECS data between
1.7--10.0 keV where the instrument responses are well determined.
The HPGSPC data were rebinned using standard binnings in the bands 7--34~keV and
PDS data were grouped appropriately for this source in which the
data decrease rapidly with energy. PDS data outside the band 150--220 keV
were ignored. 
\begin{table*}
\caption[ ]{Spectral fitting results obtained by fitting the \sax\
NFI non-dip spectrum. \nh\ is in units of $\rm {10^{22}}$ atom $\rm {cm^{-2}}$
and 90\% confidence limits are given}
\begin{flushleft}
\begin{tabular}{llllll}
\hline\noalign{\smallskip}
Model & \hfil N$_{\rm {H}}$ \hfil & kT (keV) &\hfil $\Gamma$ \hfil 
& $\rm{E_{co}}$ (keV) & $\chi^2$/dof \\
\noalign{\smallskip\hrule\smallskip}
Power law & $12.87\pm 0.2$ & \dots & $2.89\pm 0.04$ & \dots &867/297  \\
Bremsstrahlung & $9.9\pm 0.2$ & $4.73\pm 0.12$ & \dots & \dots &402/296\\
Blackbody	& $5.5\pm 0.1$ &$1.41\pm 0.01$ & \dots &\dots & 651/297\\
Disk blackbody  & $8.0\pm 0.1$ &$2.05\pm 0.03$ &\dots &\dots & 390/297\\
Cutoff power-law & $8.6\pm 0.4$ & \dots &$ 0.46\pm 0.21$ &$2.9\pm 0.3$ &396/296\\
Disk blackbody + cut-off power law &$7.2\pm 0.6$ &$5.1\pm 1.3$ &$-0.78\pm 0.57$ &$1.7\pm 0.3$ &291/294\\
Blackbody + cut-off power law & $8.6\pm 1.0$ & $1.31\pm 0.07$&$\rm {2.0^{+0.5}_{-0.8}}$& $\rm {12^{+14}_{-5}}$ &287/294 \\
\noalign{\smallskip}
\hline
\end{tabular}
\end{flushleft}
\label{tab:spec_paras}
\end{table*}
\begin{figure*}
\epsfxsize=110 mm
\begin{center}
\leavevmode
\epsffile{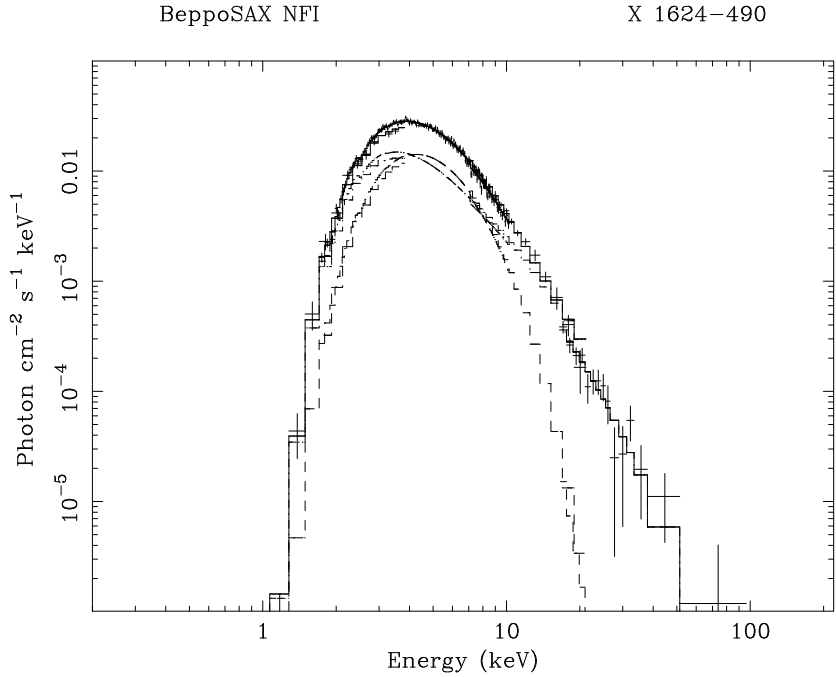}
\end{center}
\caption[]{The non-dip NFI \src\ spectrum fitted with the
absorbed cut-off power law and blackbody model discussed in the
text. The total model and the contributions of the blackbody
and cut-off power law components are
shown separately}\label{fig:m2}
\end{figure*}
In the following, the photoelectric absorption
cross sections of Morrison \& McCammon (1983) were used incorporating the
Solar abundances of Anders \& Grevesse (1989). 

Initially, the non-dip spectrum was fitted with simple models, including absorbed power-law,
thermal bremsstrahlung, blackbody and cut-off power law models.
Factors were included in the spectral fitting to allow for normalization
uncertainties between the instruments. Results are shown in Table 1.
The power-law model was completely incapable of fitting the broad-band spectrum
producing a $\chi^2$/dof of 867/297; this was only achieved at the expense 
of a PDS normalization factor of 0.28 relative to MECS which cannot be real
\begin{table*}
\caption[]{Best fits to the dip spectra. $\rm {N_H}$ is in units of $\rm {10^{22}}$ 
atom $\rm {cm^{-2}}$ for both the spectral components. 90\% confidence limits are given}
\begin{flushleft}
\begin{tabular}{lrlllr}
\noalign{\hrule\smallskip}
Spectrum &MECS ct &$\rm {N_H^{BB}}$ &$\rm {N_H^{CPL}}$ &\hfil f \hfil&
$\chi^2$/dof\\
&rate (s$^{-1}$) \\
\hline
(a) without halo components\\
\noalign{\smallskip\hrule\smallskip}
Non-dip     &9.7--10.3 &8.6$\pm $0.4 &8.6$\pm $0.4   &0.0&150/151 \\
Shallow dip &6.0--8.0  &27$\pm $7    &25$\pm $14     &$\sim $0.085    &37/28 \\
Medium dip  &4.0--6.0  &52$\pm $17   &50$\pm $30     &$\rm{0.317^{+0.047}_{-0.091}}$&24/27 \\
Deep dip    &2.0--4.0  &94$\pm $20   &63$^{+80}_{-25}$ &$\rm{0.492^{+0.119}_{-0.064}}$&2/7 \\
\noalign{\smallskip}
\hline
(b) with halo components\\
\noalign{\smallskip\hrule\smallskip}
Non-dip     &9.7--10.3 &8.6$\pm $0.4 &8.6$\pm $0.4  &0.0   &151/152 \\
Shallow dip &6.0--8.0  &25$\pm $10   &21$^{+43}_{-12}$     &0.355$\pm$0.205 &38/28 \\
Medium dip  &4.0--6.0  &45$\pm$18    &40$^{+44}_{-20}$     &$\rm{0.607^{+0.165}_{-0.089}}$ &23/27 \\
Deep dip    &2.0--4.0  &100$^{+135}_{-20}$ &40$^{+5}_{-19}$     &$\rm{0.817^{+0.003}_{-0.295}}$&2/7 \\
\noalign{\smallskip}
\hline
\end{tabular}
\end{flushleft}
\label{tab:dips}
\end{table*}
and strong discrepancies between model and data in the PDS band due to down-curving 
in the spectrum. The bremsstrahlung model similarly did not provide an acceptable fit
with a $\chi^2$/dof of 402/296. An absorbed blackbody model gave a fit with 
$\chi^2$/dof of 651/297, the spectrum being much broader than any simple
blackbody. An absorbed multi-temperature disk blackbody similarly
could not fit, the model falling below the data above 10 keV. 
An absorbed cut-off power law model also gave a poor fit
with a $\chi^2$/dof of 396/296 and a low value of the Comptonization cut-off energy
of $\rm {2.9\pm 0.3 }$ keV which is not consistent with the spectrum
extending to high energies as observed. A two-component model
consisting of a disk blackbody plus a cut-off power law gave $\chi^2$/dof
almost as good as the best model (below) but with unphysical power law
index and an inner radius for the disk blackbody of $\sim $0.4 km
which is also unphysical.

We next tried the two-component model used extensively to fit other members of
the dipping class (see Sect. 1), in which the emission regions are point-like blackbody
from the neutron star and extended Comptonized emission from the ADC.
This model gave an acceptable fit with a $\chi^2$/dof of 287/294. The
unfolded spectrum of this best-fit model is shown in Fig.~6. The values of the
normalization factors for the individual instruments (relative to the MECS)
were similar to those found for other sources.

\subsection {Spectral evolution in dipping}

After initial trials, MECS dip spectra were extracted with count rates
between 2.0--4.0 count s$^{-1}$, 4.0--6.0 count s$^{-1}$ and 6.0--8.0 count
s$^{-1}$ after first binning the data in 8 s intervals. Additionally, 
a spectrum was initially selected with 8.0--9.0 count s$^{-1}$, but this
lay too close to the non-dip spectrum for parameters to be well-determined
as is often found. Given the count rate of the source, a larger number of 
dip spectra would not be sensible. Tests showed that in this source, in
which there is fast variability during dipping,
binning the data into bins longer than 8 s before spectra were extracted
resulted in unacceptable averaging with the consequence that the
depth of dipping can appear less than it actually is. 
An acceptable spectral model must be able
to fit both non-dip and all dip spectra, without any 
changes in the parameters that characterize the source emission such as 
blackbody temperature or power law index. In the case of weak sources, it
may be necessary to fit non-dip and dip spectra simultaneously, however 
in \src\, good fits were obtained by applying the non-dip solution to the 
dip spectra and using the Progressive Covering model. This model was
firstly applied without any additional halo component.
In this model,
the emission components of the best non-dip fit are progressively covered by
absorber, i.e. the extended Comptonized emission is progressively covered
and the point-like blackbody is rapidly removed when the envelope
of the absorber overlaps the blackbody. The model flux may be written:
\[
{\rm e^{-\sigma_{MM}N_H} \;( I_{BB}  e^{-\sigma_{MM}
N_H^{BB}} \rm +\;  I_{CPL}\,(f\, e^{-\sigma_{MM}N_H^{CPL}}+\; (1 - f))}
\]

Good fits were obtained for all levels of dipping with this model as shown 
in Table 2a. It can be seen however, that the covering fraction does not
rise above 50\% which is inconsistent with
other sources in which the progressive covering fraction rises to 100\%,
e.g. XB\th 1916-053 and XBT\th 0748-676 (Church et al. 1997, 1998a,b).

\begin{figure*}
\epsfxsize=160 mm
\begin{center}
\leavevmode
\epsffile{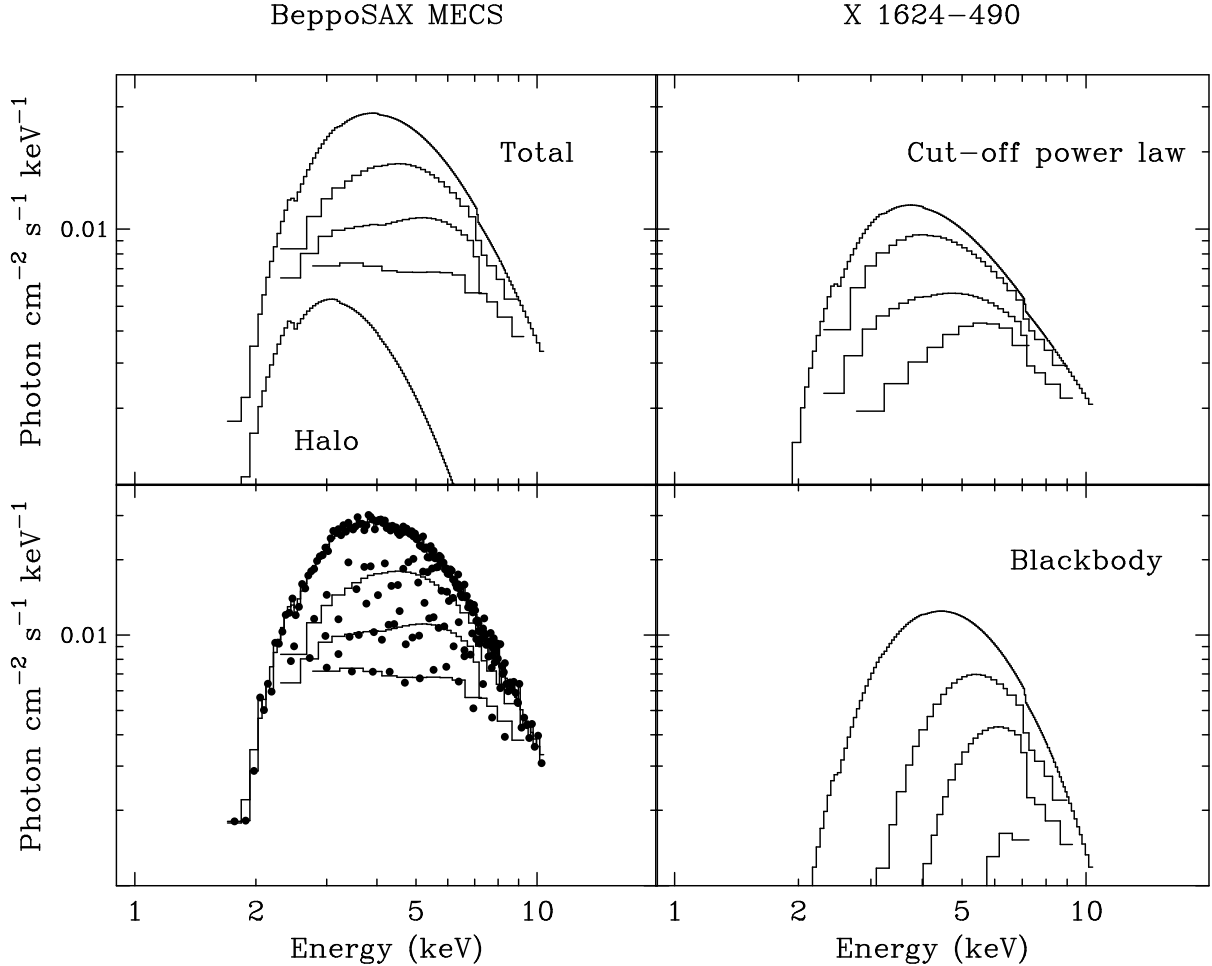}
\end{center}
\caption{Best-fit spectral fitting results to MECS non-dip and three dip
spectra. Lower panel (left) shows the data and total model; lower right
shows the blackbody; upper right shows the Comptonized component subject
to Progressive Covering; upper left shows the total model and also the
constant halo component (see text)}
\label{}
\end{figure*}

Next, we repeated this dip fitting with dust scattering terms
added to the above best-fit models, so that
\[\rm { I\; =\; I_d\;e^{-\tau}\; +\; I_n\; (1\; -\; e^{-\tau})}\]
where $\rm {I_d}$ is the source dip intensity and $\rm {I_n}$ is the
non-dip intensity. The non-dip intensity consists simply of a
blackbody plus a cut-off power law with Galactic absorption, and
$\rm {I_d}$ is given by the equation above in which these emission
components are subjected to Progressive Covering. The net model can
be expressed in the simplified form: 
$\rm {AG\,e^{-\tau}(AB.BB + PCF.CPL) + 
AG(1 - e^{-\tau})(BB + CPL)}$ 
where AG represents Galactic absorption, BB is the blackbody with
absorption AB, CPL the
cut-off power law and PCF the progressive covering fraction.
Thus, the dip intensity is reduced by the cumulative effect of
dust scattering and photoelectric absorption at every level of dipping, 
whereas the intensity scattered towards the observer is constant
depending only on the non-dip source level. The optical depth was set
to the best value of 2.4 obtained from the radial fitting, and the
energy dependence of $\tau $ set to $\rm {E^{-2}}$. Good fits were obtained to
the non-dip and dip MECS spectra and the results are shown in Table 2b
and Fig. 7. Although there is no significant improvement in $\chi
^2$/dof compared with the model without dust scattering, the results
including the effects of dust scattering are clearly preferable.
It can be seen from
Table 2b that the progressive covering factor rises to 82\% in the
deep dip spectrum (2.0--4.0 count s$^{-1}$), i.e. still not reaching
100\% which must be due to incomplete absorption of the emission
components.

In Fig.~7 the fits to the non-dip and 3 levels of dipping are shown, with the evolution 
of the blackbody and cut-off power law components shown separately. In the top left panel 
the total model including all components is shown together with the total halo component
itself containing blackbody and cut-off power law contributions from the incident spectrum 
to scattering. Firstly, it can be seen that the blackbody emission (lower right) is rapidly 
absorbed when the point-source becomes covered by the absorber, and is close to
zero flux in the deep dip spectrum. The cut-off power law clearly displays progressive 
covering in that, in each spectrum, part of the spectrum at low energies is clearly 
unabsorbed corresponding to the uncovered part of an extended source,
i.e. the accretion disk corona. The contribution of this unabsorbed part decreases 
as dipping deepens. In non-dip and shallow dipping this uncovered component dominates the spectrum
below 5 keV. In medium dipping it is about equal to the halo
contribution (at $\sim $3 keV), and in deep dipping the halo is larger
than the uncovered emission below about 4.5 keV. However, in addition
to emission uncovered by the absorber envelope, the blobbiness of the absorber demonstrated
by the fast variability in dipping will also prevent dipping reaching
100\%. Between the blobs, a fraction of the Comptonized emission from
the ADC will be transmitted.

These results are consistent with the results obtained from
{\it Exosat} by CBC95, who found in dipping there was strong absorption
of the blackbody with weaker absorption of the
Comptonized emission. Having available a set of dip spectra in the
present work, not just one dip spectrum, the existence of the
unabsorbed emission is revealed and the fact that progressive
covering takes place. 

\section {Discussion}

We have shown that spectral evolution in dipping in \src\ is well-described 
by Progressive Covering of a two-component model which assumes that X-ray 
emission consists of point-like blackbody emission from the neutron star plus
extended Comptonized emission from an ADC. The 1--30 keV luminosity of
the source is $\rm {7.3\times 10^{37}}$ erg s$^{-1}$, for a distance
of 15$\pm $5 kpc (Christian \& Swank 1997), a substantial
fraction of the Eddington limit, and the source is the most luminous
of the dipping LMXB.

The bolometric blackbody luminosity is
$\rm {2.59\times 10^{37}}$ erg s$^{-1}$, corresponding to a blackbody
radius $\rm {R_{BB}}$ of 8.3 km. If it is assumed that the emission region 
is an equatorial region on the neutron star in the plane of the accretion
disk (a sphere intersected by two planes), 
the half-height of the region {\it h} calculated using the expression
$\rm {4\, \pi\, R_{BB}^2\; =\;4\, \pi \, R_x\, h}$, where
$\rm {R_x}$ is the radius of the neutron star (assumed to be 10 km), is 6.8$\pm 1.8$ km. Thus 
the blackbody emission is from a large fraction of the stellar surface.
The half-height {\it H} of the radiatively-supported inner disk calculated
from the luminosity of the source is 6.3$\pm $2.9 km, in agreement
with the half-height of the emission region. These results were included
in a larger sample of 14 LMXB investigated using {\it ASCA} and \sax\
(Church \& Ba\l uci\'nska-Church 2000),
in which approximate agreement is found between {\it h} and {\it H}
over more than 3 decades in either parameter. This provides strong
evidence that the  neutron star and not the accretion disk is the 
source of blackbody emission in LMXB. Our spectral fitting supports
the point-like nature of the blackbody emission and the extended
nature of the Comptonized emission. The source joins the group of
other dipping sources having an unabsorbed component in dipping which
are well described by this combination of emission model with a
progressive covering absorption model 
(Church et al. 1997, 1998a,b; Ba\l uci\'nska-Church et al.
1999). The only sources not tested with this model were X\th 1746-371
and XB\th 1254-690, but it has been shown that {\it ASCA} data
on these two sources are well fitted by the model (Church et al. in preparation).
It thus appears that {\it all} of the dipping LMXB may be
explained by this combination of emission model and description of
covering during dips.

\src\ is complicated by the dust-scattered halo. Adding halo terms to
the spectral model for dip spectra showed that the covering fraction
for the extended hard emission rose to $\sim $82\%
while modelling without the halo gave a maximum fraction
less than 50\%. Spectral modelling with halo terms also showed that
in shallow and medium
dipping the unabsorbed Comptonized emission dominates the low energy
spectrum, but in deeper dipping the halo becomes dominant at low
energies. In the deepest dip spectrum, the halo exceeds the uncovered
Comptonized emission below 4.5 keV. Angelini et al. (1997)
in an analysis of {\it ASCA} data also used the X-ray image to
estimate the extent of the halo, and suggested that the halo 
could explain the soft excess in dip spectra. The present work
shows that the halo dominates in deepest dipping, but in less deep
dipping the uncovered Comptonized emission of the ADC 
is the major origin of the soft excess. 

We have modelled the
radial intensity distribution function in several energy bands to
obtain the halo fraction in these bands, and also the optical depth to
dust scattering at 1 keV which was found to be $\tau $ = 2.4$\pm $0.4.
This value is larger than for any of the sources studied by Predehl \&
Schmitt (PS95), and the halo has appreciable effect on the 
the dip spectra. It is not sensible to regard
dust parameters obtained from the modelling as providing definitive
information on interstellar dust, since the radial modelling does
not depend sensitively on the dust model used. 

Dipping allows us to estimate the size of the ADC from dip ingress
times. The fast variability on timescales of \hbox{$\sim $32 s} is clearly
associated with individual blobs of absorber covering the point-source
blackbody and so is unrelated to the size of the ADC. However, other
observations of the source have consistently shown extended, relatively
shallow shoulders to the deep dipping which we can identify with the
extended emission region being progressively covered by the absorber.
This is just detectable in Fig. 1.
Using the {\it Exosat} observation and the {\it RXTE} observation of
Sept. 1999 (Smale et al. in preparation), we estimate an average
ingress time for this slow modulation of 12500$\pm $2000 s. The
dip (shoulder) ingress time $\rm {\Delta t}$ is the time taken for the bulge
in the outer disk to cross the diameter of the X-ray emitting ADC, 
and is obtained via the velocity of the outer disk given by
\[\rm {2\,\pi\, r_{disk}/P \;=\; d_{ADC}/\Delta t,}\]
\noindent
where {\it P} is the orbital period. Using a period of 20.87 hr and a mass of the
neutron star of 1.4$\rm {M_{\sun}}$ we derive an accretion disk radius
(Frank et al. 1987) of $\rm {1.0\times 10^{11}}$ cm and thus a radius of
the ADC of $\rm {5.3\times 10^{10}}$ cm. It is remarkable that in the
case of this source, the ADC extends to $\sim $50\% of the accretion
disk radius, compared with $\sim $10\% in XB\th 1916-053 and XB\th 1323-619  
(Church et al. 1998b; Ba\l uci\'nska-Church et al. 1999). However, given that
the luminosity is $\approxgt$ 10 times higher in \src\, it is not
surprising that the ADC, probably formed by evaporation by the central
source and the ADC itself, extends to much larger radii.

In summary, we have completed an X-ray study of \src\ using {\it
BeppoSAX}. We have obtained an improved orbital period of $\rm {20.87\pm 0.01}$ hr
and made the first discovery of interdipping. We have investigated the 
depth of dipping in several energy bands and carried out modelling of the radial
intensity distribution to obtain the halo fraction at several
energies. From these data, the optical depth to scattering was derived, 
and an approximate $\rm {E^{-2}}$ dependence of the optical depth confirmed. 
Spectral fitting results for the broadband non-dip spectrum are presented, and 
including a halo component in spectral fitting we have shown that
spectral evolution is entirely consistent with progressive covering
of blackbody emission from the neutron star and extended
Comptonized emission from the ADC.



\begin{thebibliography}{}

\bibitem[1989]{}
Anders E., Grevesse N., 1989, Geochimica et Cosmochimica Acta 53, 197

\bibitem[]{}
Angelini L., Parmar A.N., White N.E., 1997, Proc. IAU Colloquium 163,
Eds. D.T. Wickramasinghe, G. V. Bicknell, L. Ferrario, Port Douglas,
1996

\bibitem[]{}
Ba\l uci\'nska-Church M., Church M.J., Oosterbroek T., et al.,
1999, A\&A 349, 495


\bibitem[1997]{}
Boella G., Chiappetti L., Conti G., et al., 1997, A\&AS 122, 327

\bibitem[]{}
Christian D. J., Swank J. H., 1997, ApJ Suppl 109, 177


\bibitem[1995]{}
Church M.J., Ba\l uci\'nska-Church M., 1995, A\&A 300, 441

\bibitem[]{}
Church M.J., Ba\l uci\'nska-Church M., 2000, ApJ, submitted


\bibitem[1997]{}
Church M.J., Dotani T., Ba\l uci\'nska-Church M., et al., 1997, ApJ 491, 388

\bibitem[1998]{}
Church M.J., Ba\l uci\'nska-Church M., Dotani T., Asai K., 1998a, ApJ
504, 516

\bibitem[1998]{}
Church M.J., Parmar A.N., Ba\l uci\'nska-Church M., et al., 1998b, A\&A,
338, 556


\bibitem[1986]{}
Courvoisier T.J.-L., Parmar A.N., Peacock A., Pakull M., 1986,
ApJ 309, 265

\bibitem[]{}
Dickey J.M., Lockman F.J., 1990, ARA\&A 28, 215

\bibitem[]{}
Frank J., King A.R., Lasota J.-P., 1987, A\&A 178, 137

\bibitem[1976]{}
Frontera F., Costa E., Dal Fiume D., et al., 1997, A\&AS 122, 371

\bibitem[]{}
Jones M.H., Watson M.G., 1989, Proc. of 23rd ESLAB Symposium, Bologna

\bibitem[1976]{}
Manzo G., Guarrusso S., Santangelo A., et al., 1997, A\&AS 122, 341

\bibitem[1976]{}
Martin P.G., 1970, MNRAS 149, 221

\bibitem[]{}
Mauche C.W., Gorenstein P., 1986, ApJ 302, 371

\bibitem[1976]{}
Morrison D., McCammon D., 1983, ApJ 270, 119

\bibitem[1986]{}
Parmar A.N., White N.E., Giommi P., Gottwald M., 1986,
ApJ 308, 199

\bibitem[1997]{}
Parmar A.N., Martin D.D.E., Bavdaz M., et al., 1997, A\&AS 122, 309

\bibitem[1995]{}
Predehl P., Klose S., 1996, A\&A 306, 283 

\bibitem[1995]{}
Predehl P., Schmitt J.H.M.M., 1995, A\&A 293, 889 (PS95)

\bibitem[1992]{}
Smale A.P., Mukai K., Williams O.R., Jones M.H.
Corbet R.H.D., 1992, ApJ 400, 330

\bibitem[]{}
Stark A.A., Gammie C.F., Wilson R.W., et al., 1992, ApJS 79, 77

\bibitem[]{}
Watson M.G., Willingale R., King A.R., Grindlay J.E., Halpern J.,
1985, IAU Circ. 4051

\bibitem[1982]{}
White N.E., Swank J.H., 1982, ApJ 253, L61

\end{thebibliography}
\end{document}